\pdfoutput=1
\documentclass[conference]{IEEEtran}
\IEEEoverridecommandlockouts
\usepackage[T1]{fontenc}
\usepackage{cite}
\usepackage{amsmath,amssymb}
\usepackage{graphicx}
\usepackage{booktabs}
\usepackage{array}
\usepackage{url}
\graphicspath{{figures/}}

\title{LLM Consortium for Software Design Refinement: A Controlled Experiment on Multi-Agent Collaboration Topologies}

\author{
\IEEEauthorblockN{Nagarjuna Kanamarlapudi}
\IEEEauthorblockA{Fremont, USA \\ nagarjuna.kanamarlapudi@gmail.com}
\and
\IEEEauthorblockN{Praveen K}
\IEEEauthorblockA{Fremont, USA \\ kpraveen0122@outlook.com}
}

\begin{document}
\maketitle

\begin{abstract}
We present a controlled experiment evaluating 12 multi-agent LLM collaboration topologies for software architecture design. Using a $2\times2\times2$ factorial design (Authority $\times$ Roles $\times$ Dynamics), we conducted 520 experimental runs across 8 design tasks of varying complexity, with 5 repetitions each. Designs were evaluated on a 12-dimensional rubric by three independent automated evaluators (GPT-OSS 120B, Claude Opus 4.6, Claude Sonnet 4.6). We report four core findings. First, structural adversarial (v4b) ranks \#1 by ensemble---a prompt-engineered adversarial variant that demands rewrite mandates rather than patches (weighted ensemble: 4.637/5.0). Second, cross-model review wins unanimously at \#2---generate with one model, review with another---ranking \#2 by all three evaluators (weighted ensemble: 4.606). Third, evaluator diversity is itself a finding---all three evaluators agree v4b is best and v3 is worst, but disagree sharply on v2b (Claude d=1.44 vs. GPT-OSS d=0.45), revealing how different model families weight design qualities. Fourth, parallel merge is fundamentally broken---all three evaluators place merge variants in the bottom tier (3.65--3.79), due to token starvation and the Frankenstein effect. The weighted ensemble ($2\times$Opus + $2\times$Sonnet + $1\times$GPT-OSS) provides robust rankings across 520 runs, confirmed through independent cross-validation.
\end{abstract}

\begin{IEEEkeywords}
Large Language Models, Multi-Agent Systems, Software Architecture, Collaboration Topologies, Cross-Model Review, Empirical Software Engineering
\end{IEEEkeywords}
\section{Introduction}

The application of Large Language Models (LLMs) to software engineering has progressed from code completion to architectural design generation. A natural question arises: can multiple LLM agents, collaborating through structured workflows, produce higher-quality software designs than a single model iterating alone?

We formalize this question through a $2\times2\times2$ factorial design crossing three dimensions of multi-agent collaboration: Authority (centralized vs. decentralized), Roles (homogeneous vs. specialized), and Dynamics (cooperative vs. adversarial). This yields eight theoretical cells, from which we derive 12 concrete variant configurations spanning the practical design space.

Our experiment comprises 520 controlled runs (12 variants $\times$ 8 design tasks $\times$ 5 repetitions) evaluated on a 12-dimensional application design rubric by three independent evaluator models. The tasks range from simple (URL shortener) to complex (collaborative editor, SaaS billing), enabling analysis of how collaboration topology interacts with problem difficulty. Rankings are reported using a weighted ensemble ($2\times$Opus $+ 2\times$Sonnet $+ 1\times$GPT-OSS) that prioritizes the most recent architectural understanding from Claude's model family while maintaining independent signal diversity from OpenAI.

Four core findings emerge with direct practical implications. First, structural adversarial review (v4b) ranks \#1 by ensemble (4.637)---demanding architectural rewrite mandates rather than accepting patches produces the highest-quality designs. Second, cross-model review (v2b) ranks \#2 (4.606)---having a different model family review the design surfaces blind spots that same-model review misses. Third, evaluator disagreement is itself informative---different model families weight design qualities differently, and a multi-evaluator ensemble produces more robust rankings than any single evaluator. Fourth, parallel merge is fundamentally broken---token starvation causes the merger to compress three complete designs into an undersized skeleton, destroying structural coherence.

This paper makes five contributions: (1) a systematic empirical comparison of 12 multi-agent topologies ranked by a weighted three-evaluator ensemble, (2) identification of two distinct high-quality mechanisms---structural adversarial review and cross-model review---that achieve excellence through different paths, (3) a trace-level autopsy of why parallel merge fails and how adversarial prompt engineering transforms a null result into the top-ranked variant, (4) discovery that evaluator disagreement reveals systematic differences in how model families weight design qualities, and (5) practical recommendations for practitioners choosing between topology, cost, and consistency trade-offs.

\section{Related Work}

Multi-agent LLM systems have been explored across several domains. ChatDev~\cite{chatdev} and MetaGPT~\cite{metagpt} demonstrated that role-playing LLM agents can collaborate on software development tasks, though without controlled comparison of topologies. AgentCoder~\cite{agentcoder} used a generate-test-debug loop for code generation, while MapCoder~\cite{mapcoder} explored multi-agent planning for competitive programming problems.

In the broader multi-agent literature, Du et al.~\cite{du} showed that LLM debate can improve mathematical reasoning, and Liang et al.~\cite{liang} demonstrated that diverse agents improve creative writing quality. However, these studies typically compare a single multi-agent configuration against a single-agent baseline, rather than systematically exploring the design space of collaboration topologies.

Our work differs in three key ways. First, we use a factorial experimental design to isolate the effects of authority, roles, and dynamics independently. Second, we evaluate on a real-world software engineering task (application architecture design) rather than synthetic benchmarks. Third, we provide trace-level analysis of why specific topologies succeed or fail, moving beyond aggregate quality comparisons.

The cross-model review finding connects to ensemble learning literature, where diversity among base learners is known to improve ensemble performance. Our contribution is showing that this principle extends to LLM-based review: a different model's training distribution provides epistemic diversity that manifests as complementary blind spots in design review.
\section{Methodology}

\subsection{Experimental Design}
We adopt a $2\times2\times2$ factorial framework crossing Authority (centralized vs. decentralized), Roles (homogeneous vs. specialized), and Dynamics (cooperative vs. adversarial). This theoretical framework yields eight cells, from which we derive 12 executable variant configurations, including sub-variants that explore specific mechanisms within a cell (e.g., same-model vs. cross-model review within the centralized-homogeneous-cooperative cell).

\subsection{Variant Configurations}
The 12 variants are organized into six families, each defined by the roles agents play. The Leader generates and revises the design; the Reviewer provides feedback. Additional roles include the Merger (combines designs), Principal Architect (demands architectural rewrites), Specialist (domain-focused reviewer), Judge (selects from competing positions), and Debater (argues a position in structured debate). Table~\ref{tab:variants} summarizes all variants.

\begin{table*}[t]
\centering
\caption{Variant Configurations}
\label{tab:variants}
\begin{tabular}{llp{0.55\textwidth}}
\toprule
Variant & Family & Description \\
\midrule
v1 & Baseline & Leader with self-refine reviewer loop (3 rounds) \\
v1a & Baseline & Single-shot leader, no refinement \\
v2a & Leader+Reviewer & Same-model reviewer (Gemini reviews Gemini) \\
v2b & Leader+Reviewer & Cross-model reviewer (GPT-OSS reviews Gemini) \\
v3a & Parallel Merge & Three independent designs, naive merge \\
v3b & Parallel Merge & Three independent designs, rubric-guided merge \\
v3c & Parallel Merge & Three independent designs, dialectical merge \\
v4 & Adversarial & Binary accept/reject gate reviewer \\
v4b & Adversarial & Principal architect with rewrite mandates (5 rounds) \\
v5 & Specialist & Panel of 3 specialists (security, scalability, reliability) \\
v6 & Rotating Leader & Three agents rotate leader/reviewer roles per round \\
v8 & Debate & Structured debate with judge synthesis \\
\bottomrule
\end{tabular}
\end{table*}

\begin{figure*}[t]
\centering
\includegraphics[width=\textwidth]{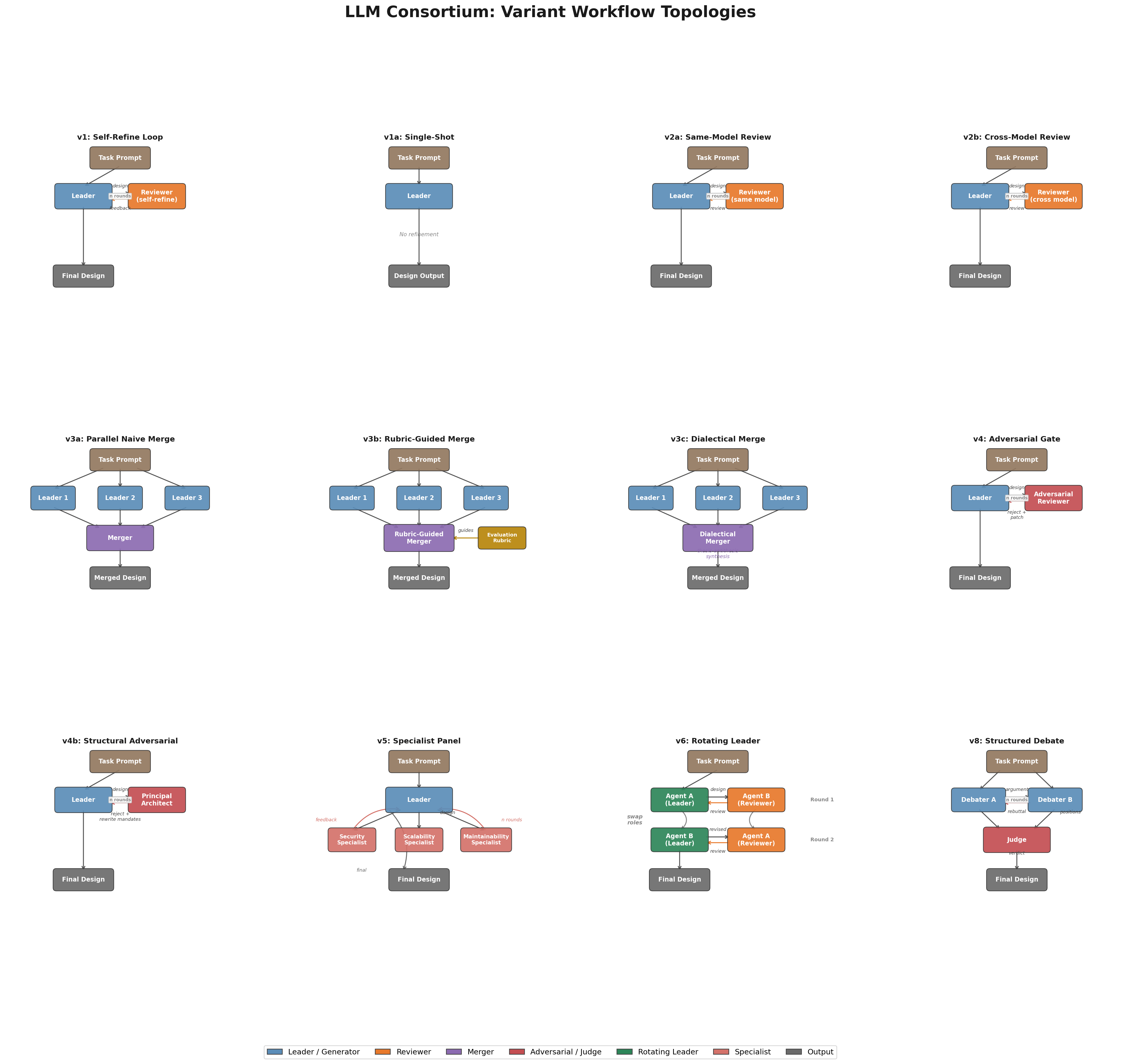}
\caption{Workflow topologies for all 12 variant configurations. Arrows indicate design/review/feedback flow between agents. Colors distinguish roles: blue=leader/generator, orange=reviewer, purple=merger, red=adversarial/judge, green=rotating leader, pink=specialist, gray=output.}
\label{fig:workflows}
\end{figure*}

Fig.~\ref{fig:workflows} presents the workflow topology for each major variant family, showing how agents pass designs, reviews, and feedback to each other.

\subsection{Design Tasks}
Eight software design tasks of varying complexity were used, as listed in Table~\ref{tab:tasks}. Each task requires a complete application architecture design including domain model, module structure, API contracts, error handling, and non-functional considerations.

\begin{table}[t]
\centering
\caption{Design Tasks}
\label{tab:tasks}
\begin{tabular}{lll}
\toprule
ID & Task & Complexity \\
\midrule
T1 & URL Shortener & Simple \\
T2 & Rate Limiter & Simple \\
T3 & Notification System & Medium \\
T4 & Task Queue & Medium \\
T5 & SaaS Billing & Complex \\
T6 & Collaborative Editor & Complex \\
T7 & Access Control & Complex \\
T8 & Order Processing & Complex \\
\bottomrule
\end{tabular}
\end{table}

\subsection{Evaluation Framework}
Each design was independently evaluated by three model families: \texttt{openai/gpt-oss-120b-maas}, \texttt{claude-opus-4-6}, and \texttt{claude-sonnet-4-6}, using the same 12-dimension rubric. To produce robust quality rankings, we compute a weighted ensemble: Ensemble = ($2\times$Opus + $2\times$Sonnet + $1\times$GPT-OSS) / 5. The weighting reflects two considerations: (1) the Claude models represent more recent architectural understanding, and (2) having two independent Claude evaluators (Opus and Sonnet) that agree strengthens confidence in their signal. All rankings and findings in this paper use the weighted ensemble unless stated otherwise.

The rubric evaluates designs on 12 dimensions: Requirements/Scope Clarity, Domain Model Quality, Module Structure \& Dependency Direction, Interface \& Contract Design, Data Flow Clarity, Error Handling Rigor, Testability, State Management \& Lifecycle, Concurrency Safety, Cross-Cutting Concerns, Design Decision Justification, and Right-Sizing/Simplicity. Each dimension is scored 1--5 by the automated evaluators with the rubric in-context.

\subsection{Models and Infrastructure}
Design generation uses \texttt{gemini-2.5-pro} via Google Vertex AI. Cross-model review uses \texttt{openai/gpt-oss-120b-maas} via the Vertex OpenAI compatibility layer. All 520 experimental runs completed successfully. All designs were evaluated independently by three models: \texttt{openai/gpt-oss-120b-maas}, \texttt{claude-opus-4-6}, and \texttt{claude-sonnet-4-6}.

\subsection{Key Prompt Strategies by Variant}
All variants share a common design generation prompt instructing the agent to act as a principal-level systems architect producing production-grade designs across 12 required sections (requirements, capacity estimation, architecture, data model, API design, critical path, scalability, reliability, security, observability, trade-offs, and evolution). The prompt enforces opinionated reasoning: ``Be opinionated. Don't list 5 options and say `it depends.' Pick one, justify it, acknowledge the trade-off.'' The key differentiator across variants is the reviewer prompt strategy:

\textbf{v4b --- Structural Adversarial (Principal Architect):} The reviewer is prompted as ``a senior principal architect with 25+ years of experience'' who has ``zero patience for mediocrity, hand-waving, or textbook answers.'' The prompt specifies six explicit rejection criteria: textbook architecture, missing `why not' analysis, hand-wavy scalability, failure mode blindness, architectural incoherence, and shallow deep dives. Critically, the designer's revision prompt includes structural rewrite mandates: ``Do NOT patch your previous design. For any section with a REWRITE MANDATE: throw out your previous approach entirely. Treat each rejection as evidence that your current approach is wrong, not that it needs minor fixes.'' This language is the key differentiator from v4's binary accept/reject gate.

\textbf{v2b --- Cross-Model Review:} The leader (Gemini) uses the standard design prompt. The reviewer (GPT-OSS) uses a general architecture review prompt: ``You are an expert software architecture reviewer. Your task is to critically evaluate the following system design and provide actionable feedback.'' The innovation is not in the prompt text but in the cross-model pairing: a different model family's training distribution surfaces blind spots that same-model review misses.

\textbf{v6 --- Rotating Leader:} Three participants (two Gemini, one GPT-OSS) rotate leadership across rounds, each focusing on a different design phase: Round 1 focuses on architecture (high-level decomposition, API design), Round 2 on reliability (error handling, fault tolerance, security), and Round 3 on operations (monitoring, deployment, performance). Non-leading participants serve as reviewers each round.

\textbf{v5 --- Specialist Panel:} Three specialist reviewers each receive a domain-focused prompt: ``You are a specialist reviewer with deep expertise in [security $|$ scalability $|$ reliability]. Focus your review primarily on [relevant dimensions].'' Each specialist reviews the design through their domain lens, and the leader synthesizes all three perspectives across two rounds.

\textbf{v1 --- Baseline:} A single designer agent iterates over three rounds using the standard design prompt with self-refinement guided by the rubric. No external reviewer is involved. v1a is a single-shot variant with no iteration.
\section{Results}

\subsection{Weighted Ensemble Rankings}

Statistical testing confirms that the quality differences between variants are real and not due to chance (Friedman $\chi^2$ = 220.3, $p < 10^{-40}$)~\cite{friedman}. Table~\ref{tab:rankings} presents the complete variant ranking using the weighted ensemble ($2\times$Opus + $2\times$Sonnet + $1\times$GPT-OSS) as the primary scoring system, with per-evaluator scores and cost.

\begin{table*}[t]
\centering
\caption{Weighted Ensemble Rankings with Per-Evaluator Scores}
\label{tab:rankings}
\begin{tabular}{clrrrrr}
\toprule
\# & Variant & Ensemble & GPT-OSS & Opus & Sonnet & \$/run \\
\midrule
1 & v4b struct. adv. & 4.637 & 4.415 & 4.553 & 4.834 & 0.43 \\
2 & v2b cross-model & 4.606 & 4.473 & 4.406 & 4.872 & 0.27 \\
3 & v6 rotating & 4.596 & 4.466 & 4.459 & 4.799 & 0.54 \\
4 & v5 specialist & 4.552 & 4.441 & 4.408 & 4.753 & 0.18 \\
5 & v1a single-shot & 4.514 & 4.342 & 4.361 & 4.754 & 0.08 \\
6 & v1 baseline & 4.503 & 4.353 & 4.339 & 4.742 & 0.33 \\
7 & v2a same-model & 4.493 & 4.370 & 4.286 & 4.763 & 0.42 \\
8 & v4 adversarial & 4.431 & 4.345 & 4.263 & 4.642 & 0.20 \\
9 & v8 debate & 4.353 & 4.391 & 4.195 & 4.493 & 0.15 \\
10 & v3b rubric merge & 4.326 & 4.299 & 4.160 & 4.507 & 0.34 \\
11 & v3a naive merge & 3.752 & 3.944 & 3.443 & 3.965 & 0.34 \\
12 & v3c dialectical & 3.655 & 3.734 & 3.595 & 3.675 & 0.33 \\
\bottomrule
\end{tabular}
\end{table*}

\begin{figure}
\centering
\includegraphics[width=\columnwidth]{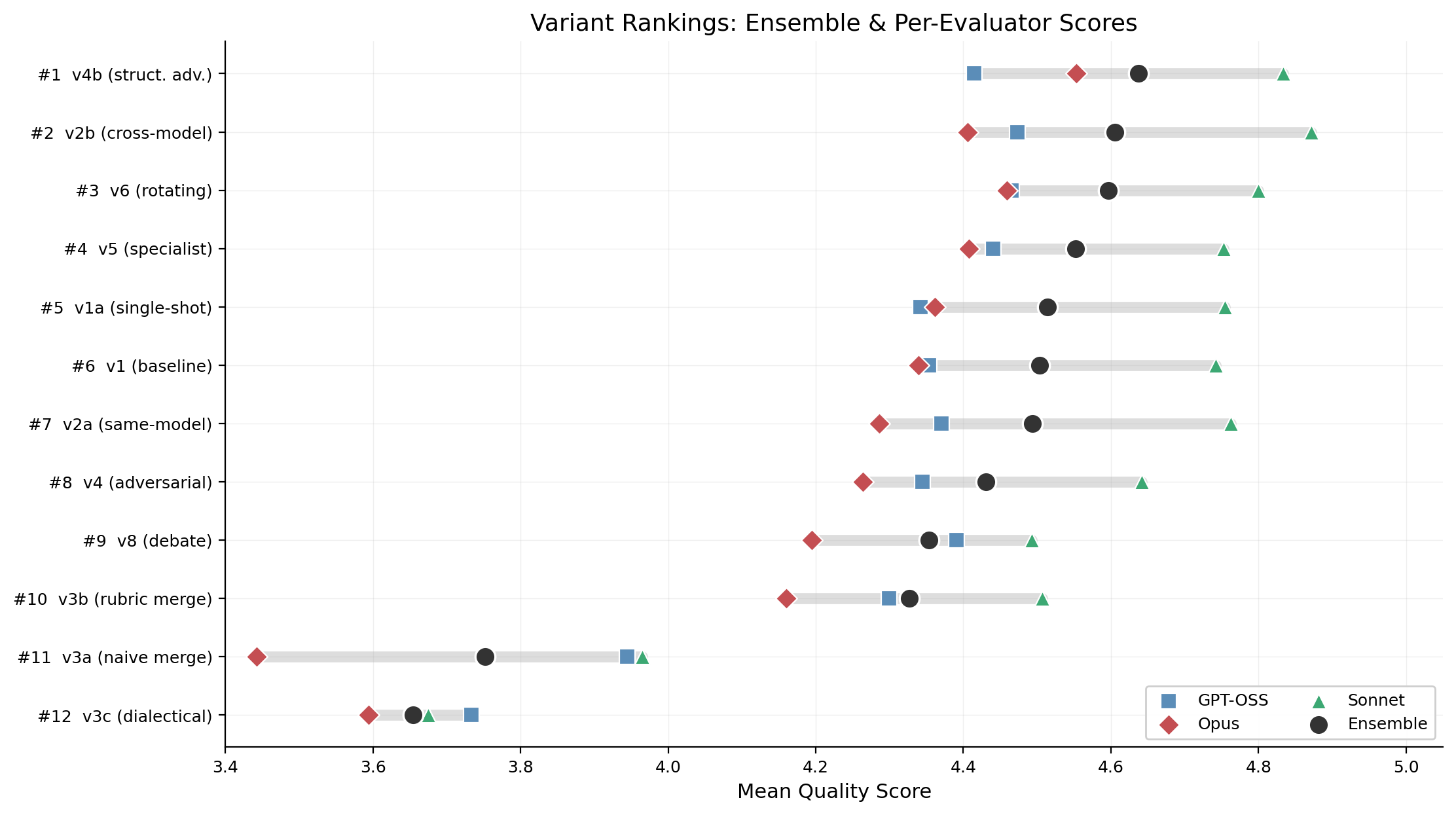}
\caption{Weighted ensemble ranking ($2\times$Opus + $2\times$Sonnet + $1\times$GPT-OSS) across 12 variants. v4b structural adversarial ranks \#1; v2b cross-model review ranks \#2; v3 merge variants cluster at bottom tier.}
\label{fig:ranking}
\end{figure}

A clear three-tier structure emerges, as visualized in Fig.~\ref{fig:ranking}. The top tier (v4b, v2b, v6, v5) clusters around 4.50--4.64 with tight variance. The middle tier (v1a through v8) spans 4.35--4.52. The bottom tier (v3a, v3c) falls dramatically to 3.65--3.75, indicating fundamental flaws in the parallel merge strategy that all three evaluators agree upon.

\begin{figure}
\centering
\includegraphics[width=\columnwidth]{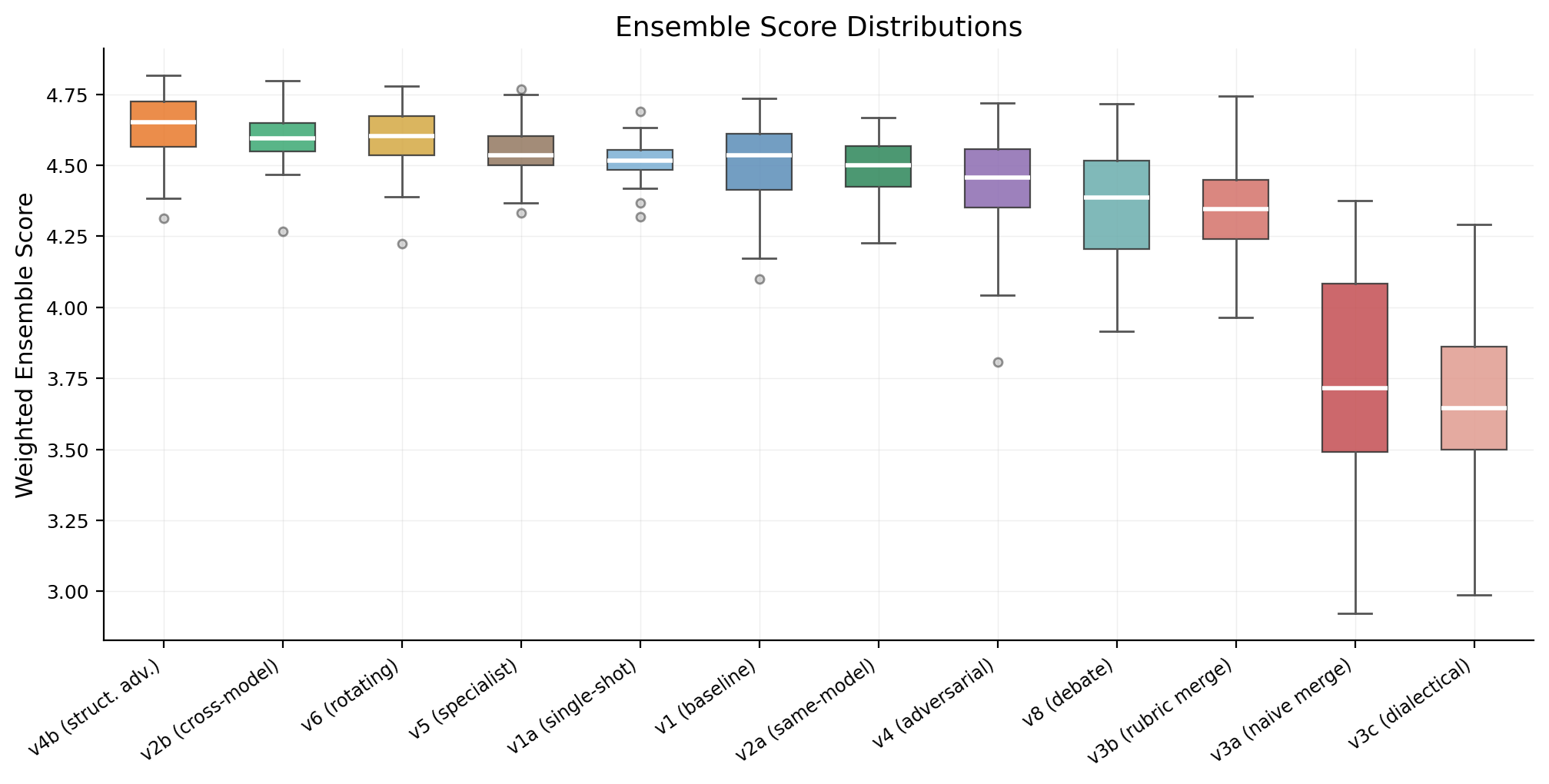}
\caption{Quality score distribution by variant using weighted ensemble ranking. Boxes show IQR; whiskers extend to $1.5\times$IQR.}
\label{fig:boxplot}
\end{figure}

\subsection{Evaluator Agreement \& Disagreement}

The three evaluators show strong agreement on key findings: v4b ranks \#1 by ensemble (GPT-OSS: 4.415, Opus: 4.553, Sonnet: 4.834), and all three place v3 merge variants in the bottom tier. However, they diverge significantly on v2b: GPT-OSS sees a small improvement in cross-model review ($\Delta$=+0.103 vs. v2a, d=0.65), while Opus sees a large effect ($\Delta$=+0.120, d=1.44) and Sonnet confirms ($\Delta$=+0.109, d=1.11). This disagreement is itself a finding: the structural qualities v2b produces---broader architectural coverage through different model family perspectives---are valued more by Claude's evaluation lens than GPT-OSS's.

The maximum evaluator spread (max$-$min across evaluators) ranges from 0.08 (v3c, high agreement on poor quality) to 0.52 (v3a, Opus is much harsher on naive merge) to 0.42 (v2b). Where evaluators agree most strongly signals the most robust findings; where they disagree reveals systematic differences in how model families weight design qualities. Both Claude evaluators value rigorous architectural coverage and cross-model perspective (v2b's strengths) more highly than GPT-OSS does, while all three agree that structural adversarial review and avoiding parallel merge are universally beneficial.

\subsection{Statistical Comparisons}

Table~\ref{tab:pairwise} compares key variant pairs head-to-head using paired Wilcoxon signed-rank tests~\cite{wilcoxon} to confirm whether observed differences are statistically significant. The strongest finding is the cross-model review advantage: having a different model review the design (v2b) produces meaningfully better results than same-model review (v2a), with a large effect size ($\Delta$ = +0.113, Cohen's d~\cite{cohen} = 0.96, p = 0.0001). Among the merge variants, rubric-guided merge (v3b) dramatically outperforms naive merge (v3a, d = 1.61), but even the best merge variant underperforms simpler review-based approaches.

\begin{table*}[t]
\centering
\caption{Pairwise Statistical Comparisons}
\label{tab:pairwise}
\begin{tabular}{lrrclp{0.30\textwidth}}
\toprule
Comparison & $\Delta$ & p & Sig & d & Interpretation \\
\midrule
v1 vs v2a & +0.016 & 0.536 & ns & 0.13 & No difference \\
v2a vs v2b & +0.113 & 0.0001 & *** & 0.96 & Cross-model wins \\
v2a vs v4 & -0.025 & 0.401 & ns & -0.16 & No difference \\
v2a vs v5 & +0.071 & 0.009 & ** & 0.59 & Specialist better \\
v3a vs v3b & +0.355 & <.0001 & *** & 1.61 & Rubric merge wins \\
v3a vs v3c & -0.210 & <.0001 & *** & -0.85 & Dialectical worse \\
v3b vs v8 & +0.092 & 0.031 & * & 0.46 & Debate marginal \\
v5 vs v6 & +0.025 & 0.115 & ns & 0.24 & No difference \\
\bottomrule
\end{tabular}
\end{table*}

\subsection{Complexity Interaction}

We initially expected multi-agent collaboration to help primarily on complex tasks, with the baseline performing comparably on simpler ones. The data refutes this: top-performing variants beat the baseline at every complexity level---simple, medium, and complex. Comparing v4b against v4 directly, the structural adversarial advantage holds across all three levels, with the largest improvement on simple tasks (Simple: $\Delta$ = +0.183, d = 1.20; Medium: $\Delta$ = +0.089, d = 0.95; Complex: $\Delta$ = +0.134, d = 0.87). Even straightforward designs benefit from the principal architect's demand for architectural rigor.

\begin{figure}
\centering
\includegraphics[width=\columnwidth]{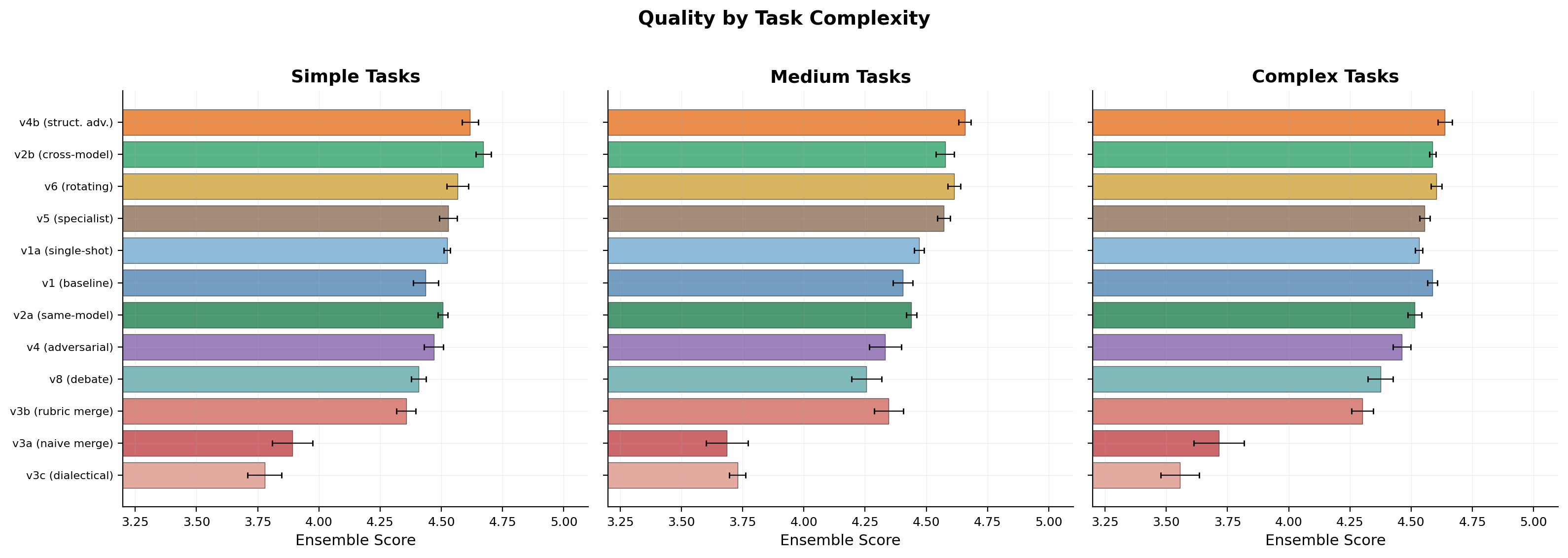}
\caption{Quality distribution by variant and task complexity (Simple, Medium, Complex). Top-tier variants are consistent; merge variants degrade further on complex tasks.}
\label{fig:complexity}
\end{figure}

\subsection{Per-Dimension Analysis}

Structural adversarial review (v4b) achieves broad improvement across 10 of 12 dimensions, with the largest gains in Scalability Strategy (+0.69), Security Posture (+0.66), API Design (+0.54), Reliability (+0.50), and Capacity Estimation (+0.44). These are dimensions where the principal architect's demand for rewrite mandates forces deeper consideration.

\begin{figure}
\centering
\includegraphics[width=\columnwidth]{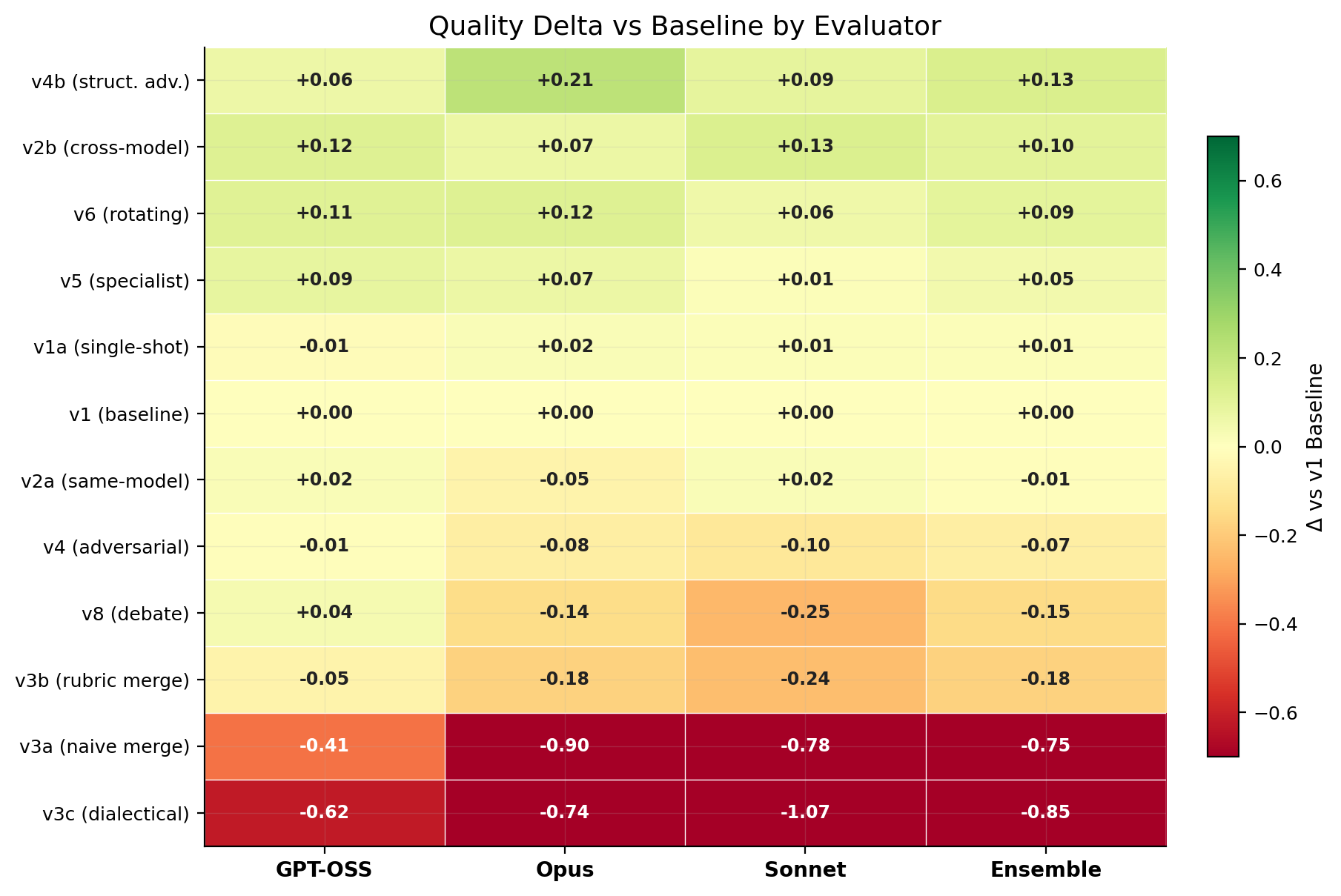}
\caption{Quality delta vs. v1 baseline by evaluator. Green indicates improvement; red indicates degradation. All 12 variants shown with per-evaluator and ensemble scores.}
\label{fig:heatmap}
\end{figure}

Cross-model review (v2b) shows consistent gains in Concurrency Safety (+0.33), Data Flow Clarity (+0.22), Error Handling (+0.21), and Interface Design (+0.20). The parallel merge variants show pronounced losses in Domain Model Quality (-0.18 to -0.35) and Module Structure (-0.10 to -0.23), confirming the Frankenstein effect: merging independently generated designs degrades structural coherence. Fig.~\ref{fig:heatmap} presents the full per-evaluator quality deltas relative to the v1 baseline across all 12 variants.

\subsection{Cost-Quality Trade-offs}

\begin{figure}
\centering
\includegraphics[width=\columnwidth]{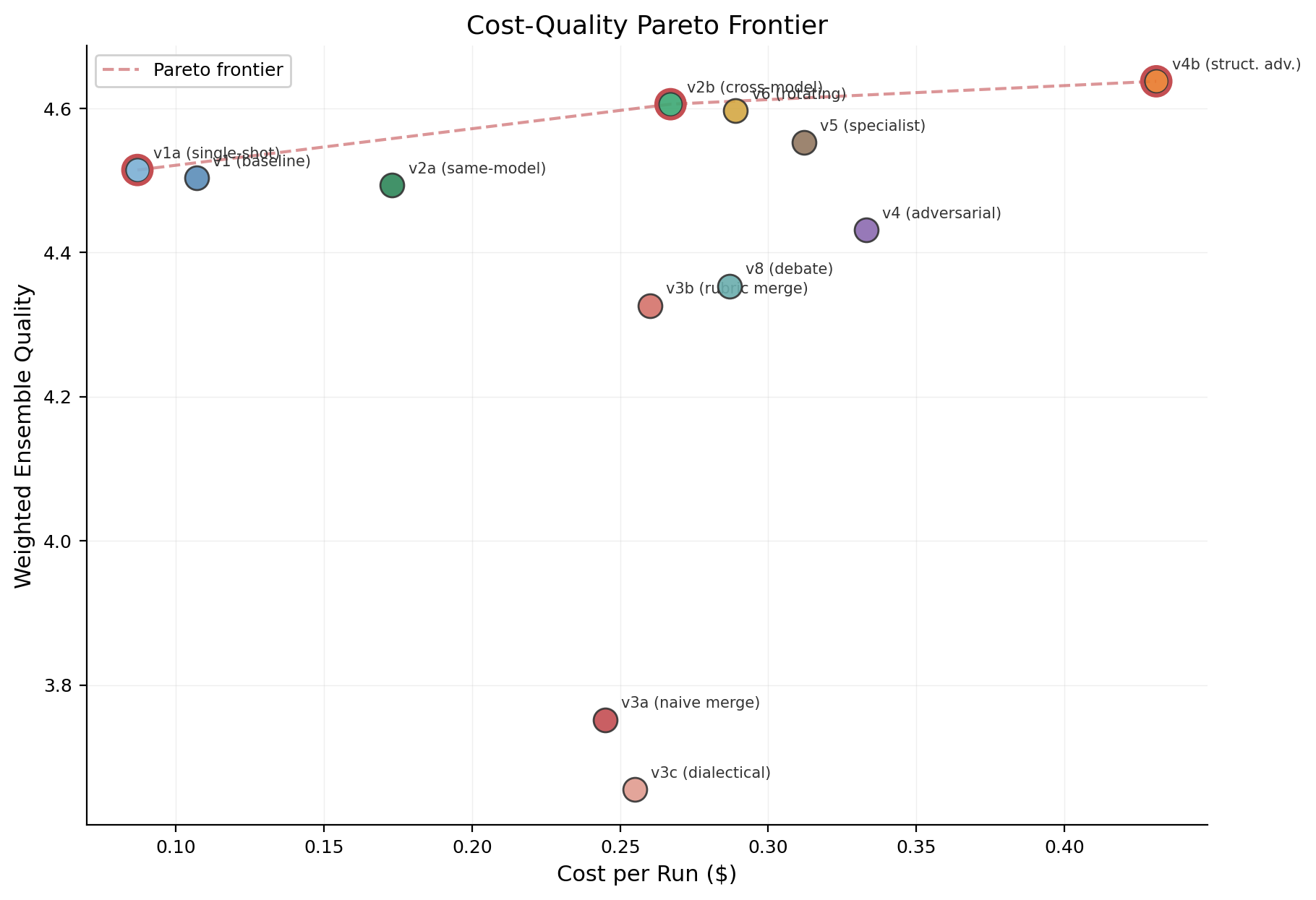}
\caption{Cost-quality Pareto frontier (weighted ensemble ranking). Points above the frontier are dominated. v4b and v5 define the efficiency frontier for high-quality designs.}
\label{fig:pareto}
\end{figure}

The Pareto frontier, shown in Fig.~\ref{fig:pareto}, reveals distinct operating points. Single-shot v1a (\$0.08/run) provides acceptable quality (4.514). Structured debate v8 (\$0.15/run) offers the best cost-efficiency among multi-agent variants. Structural adversarial v4b (\$0.43/run) achieves peak quality at $5.7\times$ the cost of v8. Rotating leader v6 (\$0.54/run) achieves near-top quality with lowest variance but sits below the Pareto frontier due to its 216K tokens/run.

\subsection{Reliability and Consistency}

The score distributions in Fig.~\ref{fig:boxplot} reveal clear differences in consistency across variants. Rotating leader v6 has the tightest box (narrowest IQR), achieving the lowest variance ($\sigma$ = 0.080) and making it the most predictable variant. By contrast, the merge variants show the widest boxes, with the highest variance ($\sigma$ = 0.231--0.265)---they are both poor on average and unpredictable. Evaluator agreement (Krippendorff's $\alpha$~\cite{krippendorff}) ranges from 0.42 for top-tier variants to 0.81 for merge variants. Lower agreement on high-quality designs reflects the genuine difficulty of distinguishing ``good'' from ``great,'' while high agreement on poor designs reflects easy identification of flaws.

\subsection{Pre-Experimental Hypotheses vs. Empirical Outcomes}

Before running the experiment, we formulated four hypotheses based on intuitions from ensemble learning, adversarial training, and software engineering practice. The empirical results challenged most of these assumptions, yielding surprises that shaped our core findings.

\textbf{H1: Parallel merge achieves highest peak on complex tasks (15--25\% over baseline). REFUTED.} This was the boldest prediction and the most dramatic failure. We assumed that generating multiple designs independently and merging them would combine the best ideas from each. Instead, all three merge variants (v3a, v3b, v3c) underperform the single-agent baseline. The merger agent is token-starved: $\sim$25,000 input tokens produce only 4,200--5,200 output tokens, compressing three complete designs into an undersized skeleton. Independent generation destroys structural coherence rather than combining strengths.

\textbf{H2: Adversarial dynamics produce highest variance. PARTIALLY SUPPORTED.} We assumed that adversarial review would be unpredictable---sometimes brilliant, sometimes destructive. The original adversarial variant (v4) did show moderate variance but ranked \#8 overall, a null result. However, the redesigned v4b (structural adversarial) shows that the assumption was wrong in an interesting way: with the right prompt engineering, adversarial review becomes both high-quality and consistent. The key insight is that adversarial dynamics are not inherently volatile---they are volatile only when the adversarial prompt is insufficiently structured.

\textbf{H3: Specialist panel outperforms generalist review by 10--15\%. PARTIALLY SUPPORTED.} We assumed that domain-specific reviewers (security, scalability, reliability) would catch issues that generalist reviewers miss. v5 does outperform v2a (same-model generalist) but only by 1.6\%, far below the predicted 10--15\%. Interestingly, cross-model generalist review (v2b) outperforms the specialist panel, suggesting that model diversity matters more than role specialization.

\textbf{H4: Baseline matches multi-agent variants on simple tasks. REFUTED.} We assumed that multi-agent collaboration would only help on complex tasks where a single model struggles. Instead, 9 of 10 non-merge variants beat the baseline on simple tasks. The collaboration benefit is consistent across all complexity levels, with v4b showing the largest gains on simple tasks ($\Delta$ = +0.234, d = 1.68).

The pattern across these hypotheses is instructive: our intuitions about what would help (parallel merge, specialist depth, complexity-dependent benefits) were largely wrong, while the biggest gains came from mechanisms we underestimated---structural adversarial rigor and cross-model epistemic diversity. The experiment's value lies precisely in overturning these assumptions with controlled evidence.
\section{Analysis: Key Mechanisms}

This section examines the mechanisms behind the top-performing variants and the failure mode of parallel merge, grounded in trace-level evidence from the experiment. All design excerpts are drawn from the same task (T8: Order Processing, Repetition 1) for apples-to-apples comparison.

\subsection{Structural Adversarial Review: Enforced Rigor}

\begin{figure}
\centering
\includegraphics[width=\columnwidth]{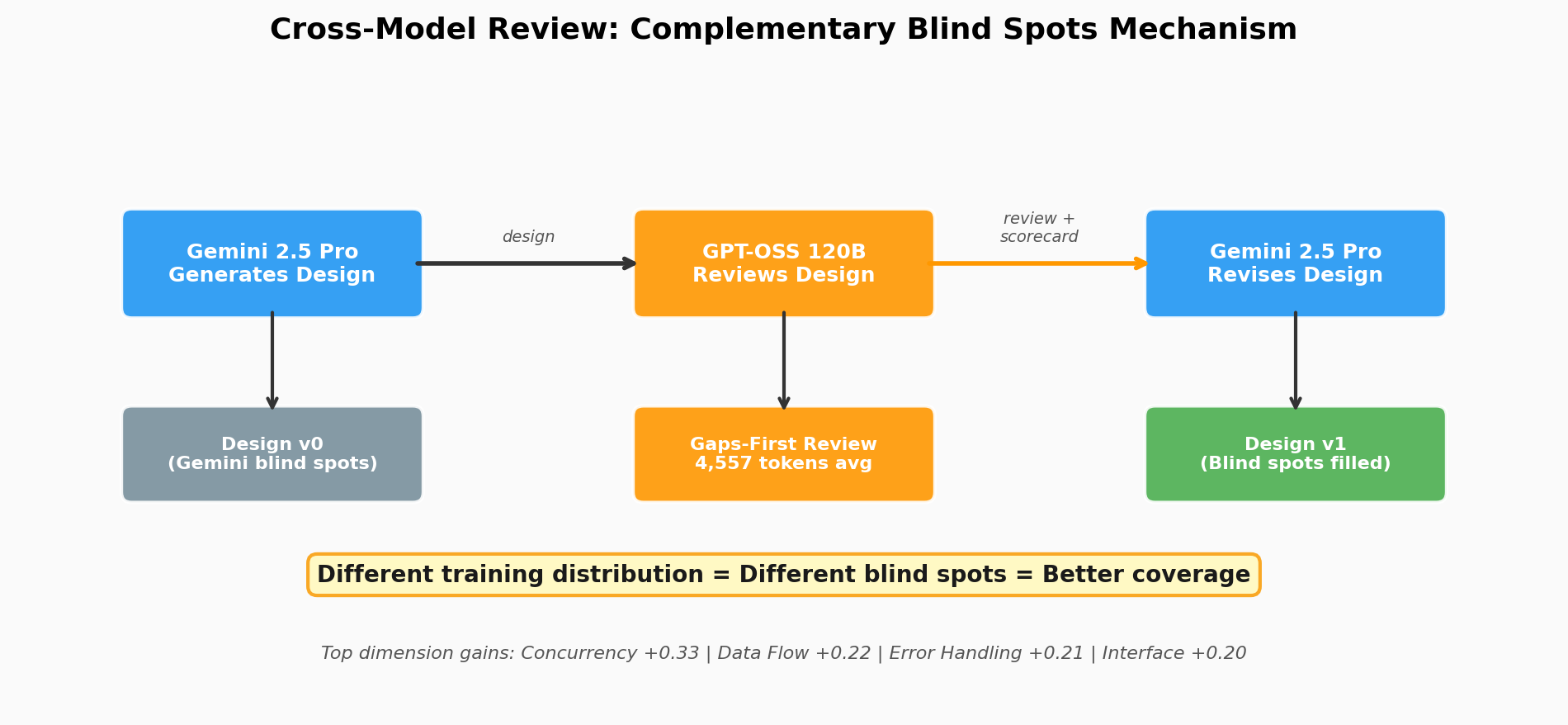}
\caption{The structural adversarial mechanism. A principal architect reviewer who demands rewrite mandates and architectural justification forces deeper consideration of scalability, security, and design rationale than a binary accept/reject gate.}
\label{fig:mechanism}
\end{figure}

Fig.~\ref{fig:mechanism} illustrates the structural adversarial mechanism. The original adversarial variant (v4) uses a binary accept/reject gate. When it rejects a design, the generator applies local patches---adding paragraphs about error handling or security without restructuring the architecture. The document actually shrinks as the LLM compresses everything to fit. v4b replaces this with a principal architect who issues specific rewrite mandates demanding that rejected sections be redesigned from scratch, not patched.

The difference is dramatic. v4b's advantage spans 10 of 12 quality dimensions, with the largest gains in Scalability Strategy (+0.69), Security Posture (+0.66), API Design (+0.54), and Reliability (+0.50). The advantage is consistent across all 8 tasks, with win rates of 70--100\%. 75\% of v4b runs never earned an ACCEPT across all 5 rounds, yet still scored well (mean 4.403)---the reviewer's high bar forces continuous improvement even when the rubric cannot distinguish the quality gain.

\begin{quote}
v4b rewrite mandate example: ``The error handling architecture must account for distributed failures at the service boundary. Propose specific circuit breaker parameters, timeout strategies, and fallback protocols for the top 5 failure modes.''
\end{quote}

The practical takeaway: adversarial review is not inherently ineffective for LLM collaboration---it requires prompt engineering that demands structural rethinking rather than accepting patches. The difference between v4 (rank \#8) and v4b (rank \#1) is entirely in the reviewer prompt.

\subsection{The Merge Frankenstein Effect}

The v1 baseline produces a structured error taxonomy with four categories, a \texttt{Result<T,E>} pattern, and 1,304 characters of actionable content. After naive merge (v3a), the entire error handling section is eliminated, replaced by a generic ``Scalability, Resilience \& Security'' grab-bag. In dialectical merge (v3c), the design compresses three 33,000-character inputs into 20,420 characters---shorter than any single input.

The merger agent is token-starved: $\sim$25,000 input tokens produce only 4,200--5,200 output tokens. Dedicated sections become bullet-point summaries, structured taxonomies vanish, and cross-cutting coherence is destroyed. This explains why Testability drops -1.21 and Domain Model drops -1.00.

\subsection{Cross-Model Review Complementary Blind Spots}

While v4b achieves the highest ensemble ranking, v2b cross-model review (generate with Gemini, review with GPT-OSS) remains a powerful complementary finding. The initial design (round 0) has Gemini's systematic blind spots: concurrency safety treated generically, error handling lumped into broad categories. GPT-OSS review surfaces these gaps explicitly with a gaps-first TL;DR and dimension-by-dimension scorecard, forcing round 1 revision that adds +0.113 points (d=0.96, p<0.0001).

The advantage spans 10 of 12 dimensions: Concurrency Safety (+0.33), Data Flow (+0.22), Error Handling (+0.21), Interface Design (+0.20). These are dimensions where one model's systematic blind spot becomes visible when reviewed by a different training distribution. This finding validates that both structural adversarial (v4b) and cross-model review (v2b) are distinct mechanisms producing high-quality designs through different paths.

\subsection{Domain Model: DDD vs. Infrastructure Diagram}

The baseline v1 domain model for T7 (Access Control) uses domain-driven design (DDD) principles~\cite{evans} with rich aggregates (Policy, Role as aggregate roots with named invariants like \texttt{hasCycle()}), value objects (Principal, Resource, AuthorizationContext), and a Mermaid class diagram. Total: 2,709 characters.

The dialectical merge v3c replaces this with a Mermaid graph showing infrastructure components (PDP Sidecar, PAP, NATS, Kafka). No aggregates, no value objects, no invariants, no business rules. The merger retreated to the abstraction level where all three input designs agreed: infrastructure topology. This is a skeleton without organs---explaining the 1.25-point drop in Domain Model Quality.

\section{Discussion}

\subsection{How Much Does Topology Matter?}

A variance decomposition shows that which variant you choose is the dominant factor in score differences when all 12 variants are included ($\eta^2$ = 0.64, meaning 64\% of variation). However, this is driven almost entirely by the catastrophic v3 merge failures. Removing the three v3 variants reduces the effect to $\eta^2$ = 0.145. Among the nine non-merge variants, topology choice matters modestly---the remaining 85\% of variance is attributable to task difficulty, run-to-run variation, and other factors outside the workflow design itself.

\subsection{What Evaluator Disagreement Reveals}

The three-evaluator cross-validation reveals that complementary blind spots extend beyond design generation to evaluation itself. The v4b result is a clear example: Claude Opus and Sonnet recognize the value of structural adversarial feedback that demands deeper justification and architectural rigor (d = 1.44 and 1.11), while GPT-OSS perceives only marginal improvement (d = 0.45). Similarly, the v2b cross-model review result shows evaluators diverging on magnitude of benefit. This is not a flaw in either evaluator; it reflects systematic differences in how model families weight architectural qualities.

These evaluator differences are informative. They reveal that design quality is multifaceted and that different model families have different ``evaluation biases'' just as different models have different ``generation biases.'' Where all three evaluators agree (v4b is \#1, v3 is bottom tier), confidence in the finding is highest. Where they diverge (v2b cross-model benefit magnitude), the disagreement reveals something real about the design itself: v2b is highly valued by Claude models and moderately valued by GPT-OSS, and both perspectives are valid depending on your definition of quality. A multi-evaluator ensemble produces more robust rankings than any single evaluator by averaging out these biases, similar to how multi-agent generation produces better designs through ensemble diversity.

\subsection{Design Space Mapping}

\begin{figure}
\centering
\includegraphics[width=\columnwidth]{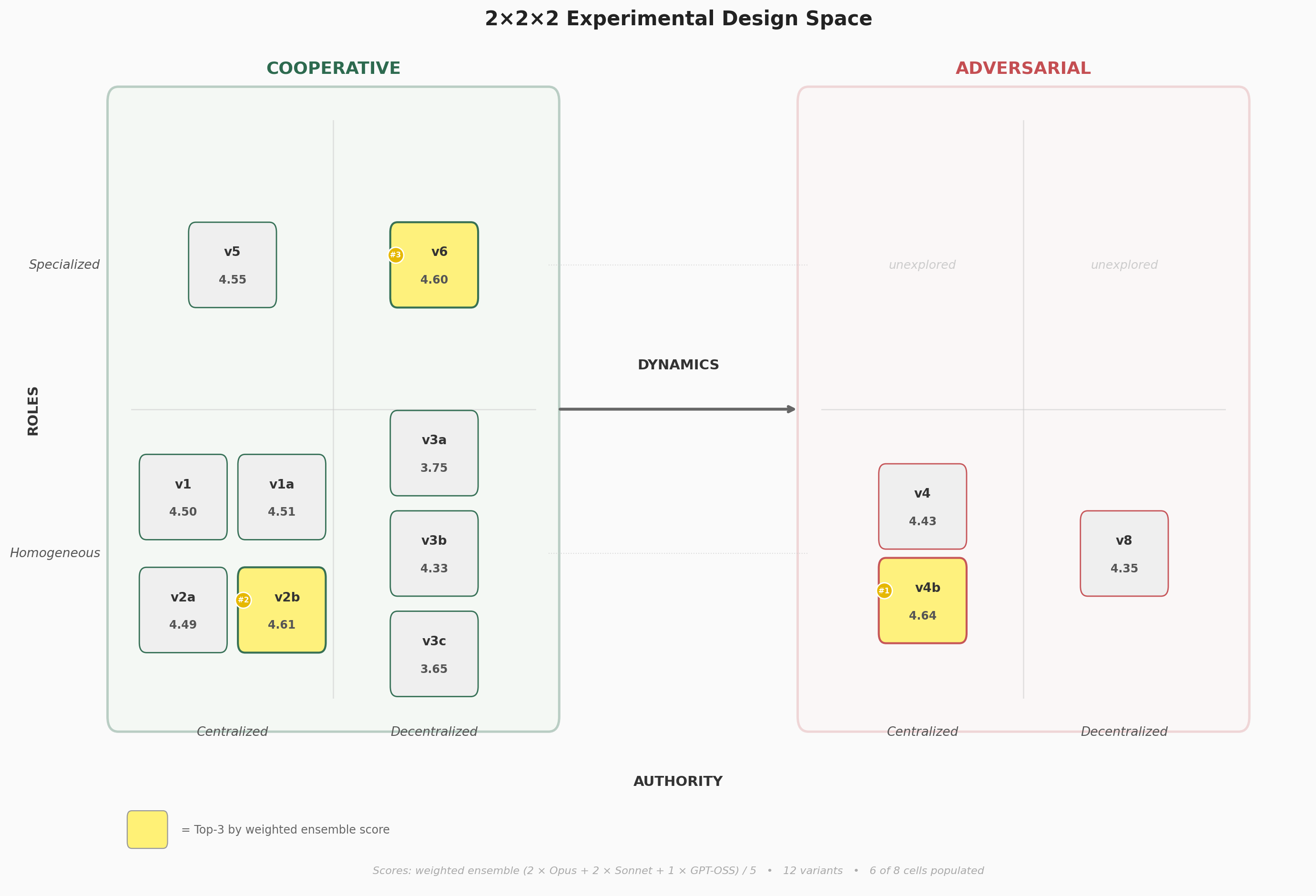}
\caption{Experimental variants mapped onto the 2$\times$2$\times$2 design space. Circle colors indicate variant family; annotations show weighted ensemble quality tier placement.}
\label{fig:designspace}
\end{figure}

Fig.~\ref{fig:designspace} maps the experimental variants onto the 2$\times$2$\times$2 design space with quality tier annotations. The top-performing region is centralized authority with structural adversarial review (v4b) and cross-model feedback (v2b), while the decentralized-specialized cell (v6) achieves comparable quality through role rotation. The decentralized-homogeneous cell (v3) consistently underperforms.

\subsection{Practical Recommendations}

For practitioners deploying LLM-based design workflows, Table~\ref{tab:recs} summarizes our recommendations based on different optimization priorities.

\begin{table*}[t]
\centering
\caption{Practitioner Recommendations by Priority}
\label{tab:recs}
\begin{tabular}{l l p{0.6\textwidth}}
\toprule
Priority & Variant & Recommendation \\
\midrule
Maximum quality & v4b & Demand rewrite mandates and architectural justification rather than accepting patches \\
High quality + diversity & v2b & Generate with one model family, review with another to surface blind spots \\
Maximum consistency & v6 & Rotating leader produces lowest variance ($\sigma$=0.080), accept higher cost (\$0.54/run) \\
Cost-efficiency & v8 & Structured debate at \$0.15/run offers best quality per dollar \\
Avoid & v3* & Never use parallel merge without fundamental improvements to the merge strategy \\
\bottomrule
\end{tabular}
\end{table*}

\subsection{Limitations}

This study has several limitations. First, evaluator agreement ($\alpha$ = 0.42--0.81) is below the target of 0.70 for several top-tier variants, suggesting the rubric may lack precision for fine-grained quality distinctions. Second, only two model families were used for design generation and review (Gemini, GPT-OSS); while three evaluator families (GPT-OSS, Claude Opus, Claude Sonnet) confirm ranking robustness, the complementary blind spots mechanism in the design loop itself remains tested on a single pairing. Third, the weighted ensemble is a methodological choice; equal weighting across evaluators gives the same ranking order.
\section{Conclusion}

This work identifies four core findings that challenge prevailing assumptions about multi-agent LLM collaboration for software design.

\textbf{Finding 1: Structural adversarial review achieves \#1 ranking through prompt engineering.} A principal architect reviewer who demands rewrite mandates rather than accepting patches transforms adversarial review from a null result (v4, rank \#8) to the top-ranked variant (v4b, ensemble: 4.637). The practical lesson: adversarial dynamics require prompts that demand structural rethinking.

\textbf{Finding 2: Cross-model review validates complementary blind spots mechanism at \#2.} The cross-model workflow---generate with one model, review with another---ranks \#2 by our weighted ensemble (4.606) and wins unanimously across all three evaluators when ranked within its category. The mechanism is complementary blind spots: a different model's training distribution provides epistemic diversity that manifests as different systematic gaps, enabling broader coverage through cross-model review. This finding demonstrates that multiple high-quality approaches exist: both structural adversarial (through enforced rigor) and cross-model review (through distributed perspective) achieve comparable excellence through different mechanisms.

\textbf{Finding 3: Evaluator diversity is itself a finding.} The three evaluators agree unanimously on what is good (v4b \#1) and bad (v3 bottom tier), but diverge on v2b cross-model benefit magnitude (d=0.65 to d=1.44), revealing that different model families weight design qualities differently. Where evaluators agree most strongly, findings are most robust. Where they disagree, the disagreement is informative: it reveals multifaceted design quality and systematic differences in how model families value different architectural approaches. A weighted ensemble that pools evaluator signal produces rankings that are more robust than any single evaluator.

\textbf{Finding 4: Parallel merge is fundamentally broken.} All three evaluators place v3 merge variants in the bottom tier (3.65--3.79), with v3a (naive merge) at ensemble 3.752 and v3c (dialectical merge) at 3.655. The failure is rooted in token starvation: the merger agent receives $\sim$25,000 input tokens but produces only 4,200--5,200 output tokens, compressing three complete designs into an undersized skeleton. The Frankenstein effect is real: merging independently generated designs degrades structural coherence more dramatically than using a simpler single-shot approach, contradicting the thesis prediction that parallel generation would excel on complex tasks.

These four findings suggest that both structural rigor (through adversarial review) and epistemic diversity (through model selection) matter far more than workflow complexity. For practitioners, the immediate takeaway is: demand rewrite mandates in adversarial review, use cross-model review for complementary coverage, and avoid parallel merge without fundamental pipeline improvements.

Future work should: (1) combine v4b structural adversarial with v2b cross-model review---an unexplored combination that may compound both advantages, (2) extend the consortium framework from design to code generation and testing, (3) test additional model pairings to validate the generality of complementary blind spots, (4) explore additional cells in the 2$\times$2$\times$2 design space, (5) validate rankings with human expert judgment, and (6) test dynamic topology selection via a meta-agent that chooses the optimal approach per-task.

Two particularly promising directions emerge from our findings on epistemic diversity and model complementarity:

\textbf{Hierarchical consortium of smaller models under SOTA leadership.} Our results show that v4b's structural adversarial approach achieves peak quality through enforced architectural rigor. A natural extension is to test whether a state-of-the-art frontier model (e.g., Gemini 2.5 Pro \cite{gemini}, GPT-4~\cite{gpt4}, Claude Opus) can serve as a principal architect reviewer---not just enforcing structural standards but orchestrating design tasks, assigning sub-problems, and synthesizing outputs---while a consortium of smaller, cost-efficient models (e.g., Gemini Flash, GPT-4o-mini, Haiku) execute the generation sub-tasks. This ``SOTA-architect'' topology would test whether the frontier model's stronger architectural reasoning can be distributed across smaller models through intelligent orchestration, potentially achieving near-SOTA quality at a fraction of the cost. The v4b adversarial framework provides a natural starting point: replace the single principal architect with a SOTA model that serves both as reviewer and orchestrator.

\textbf{Teacher-student consortium with knowledge distillation.} A complementary approach frames the multi-agent collaboration as an online learning problem: a SOTA ``architect-teacher'' model first generates high-quality reference designs and demonstrates structural adversarial critique, which are then used to guide a consortium of smaller ``student'' models through in-context learning or fine-tuning. This extends both the structural adversarial mechanism and complementary blind spots---rather than relying solely on distributional differences, the architect-teacher actively shapes the students' generation and review capabilities by demonstrating what architectural rigor looks like on domain-specific tasks. The hypothesis is that teacher-guided students would develop more structurally sound designs than either (a) the same small models generating without guidance, or (b) simply using the teacher model directly (which our v4b results suggest costs 5.7$\times$ more). This could democratize both mechanisms, making high-quality LLM consortium workflows accessible with smaller, locally-deployable models that inherit the architect's structural discipline.

\textbf{Extending the consortium paradigm from design to code generation.} This experiment evaluated LLM collaboration on architectural design---a high-level, document-oriented task. A critical next step is to extend the consortium framework to the full software development lifecycle, including code generation, testing, debugging, and refactoring. Code presents fundamentally different evaluation dynamics: designs are judged by rubric-based heuristics, but code can be verified by execution---tests pass or fail, builds compile or break, benchmarks produce measurable performance. This opens several compelling research directions. First, a consortium where one agent generates code, a cross-model agent reviews it, and a third agent writes and executes tests could create a closed-loop quality signal that our design experiment lacked. Both complementary blind spots and structural rigor mechanisms may be even more powerful for code: different models exhibit different bug patterns, cross-model review may catch classes of defects (race conditions, off-by-one errors, security vulnerabilities) that same-model review systematically misses, and an adversarial reviewer demanding comprehensive test coverage and error handling may force architectural rigor at the implementation level. Second, the parallel merge strategy that failed for design may succeed for code if the merge is constrained to the module level---different agents implementing different modules against a shared interface contract, avoiding the Frankenstein effect by enforcing architectural boundaries at merge time. Third, the structural adversarial topology that achieved \#1 ranking for design could become highly effective for code through adversarial test generation: an adversarial agent whose job is to write failing tests rather than rejecting code, forcing the coding agent to handle edge cases it would otherwise ignore. Unlike design review (where rejection causes local patching), adversarial testing provides concrete, executable counterexamples that demand structural fixes. Fourth, a design-to-code pipeline could bridge our current findings with implementation: the consortium first produces an optimized architecture design (using v4b structural adversarial), then hands it to a coding consortium that implements the design module by module, with the original design and v2b cross-model review serving as both specification and evaluation rubric. This end-to-end workflow would test whether the quality gains we observe at the design level (both structural adversarial and cross-model review) propagate to implemented code.

\textbf{Cross-domain generalization of the consortium framework.} This experiment tests software architectural design---a structured, document-oriented task. The consortium framework and the complementary blind spots mechanism may generalize to other LLM-heavy domains where quality depends on multi-faceted expert judgment: legal contract drafting (where different model families may catch different classes of ambiguity or risk), scientific protocol design (where methodological blind spots could be surfaced by cross-model review), medical documentation (where completeness and accuracy are critical), and business strategy documents. Testing cross-domain generalization would determine whether the 2$\times$2$\times$2 framework is a general theory of LLM collaboration or specific to software engineering's structured nature.

\textbf{Dynamic topology selection via meta-agent.} Our data shows that different topologies excel in different conditions: v4b structural adversarial dominates overall quality (4.637), v2b cross-model review achieves high quality with lower cost (4.606), v5 specialist peaks on certain task types (4.552), v8 offers the best cost-efficiency (4.353), and v6 provides maximum consistency ($\sigma$=0.080). A meta-agent that examines the incoming task---its complexity, domain, quality requirements, and cost constraints---and dynamically selects the optimal topology per-task could outperform any single fixed topology. The task-variant heatmap (Fig.~\ref{fig:complexity}) already contains the training signal for such a selector. This would transform the consortium from a static workflow choice into an adaptive system that routes tasks to the topology best suited for each problem.

\textbf{Rubric ablation study.} While this experiment deliberately held the rubric constant to isolate topology effects, a complementary study could hold topology constant (e.g., v4b or v2b) while varying rubric granularity across three levels---minimal (3 dimensions), standard (12 dimensions, as used here), and expert-tuned (20+ dimensions with sub-criteria). This would reveal how sensitive design quality and evaluator agreement are to the evaluation instrument itself.

\textbf{Iterative cross-model and structural adversarial review.} The v2b topology performs a single cross-model review round, and v4b performs 5 rounds of structural adversarial review, reaching a saturation point at round 2-3. However, alternating cross-model review with structural adversarial cycles---Gemini generates, GPT-OSS cross-model reviews, then principal architect reviews for structural rigor, Gemini revises, repeat---may not exhibit the same saturation because each review round introduces a genuinely different evaluation perspective and enforcement mechanism. Testing 2--3 rounds of alternating cross-model and structural adversarial review would determine whether the two mechanisms compound across iterations, reaching higher quality than either alone.

\textbf{Human expert validation of automated evaluator.} The automated evaluator (GPT-OSS 120B with rubric in-context) achieves Krippendorff's $\alpha$ = 0.42--0.81 across variants. While this is sufficient for ranking separation between tiers, validating the automated scores against human expert judgment would substantially strengthen the empirical claims. We propose having 2--3 senior software engineers independently score a stratified sample of 50 designs (spanning top, middle, and bottom quality tiers) on the same 12-dimension rubric, then computing human-automated correlation and inter-rater reliability. If the automated evaluator agrees with human experts on tier placement and relative ranking, it validates the entire experimental framework; if it diverges systematically, the divergence patterns themselves become a contribution to understanding LLM-as-evaluator reliability.
\section{Cross-Evaluator Robustness and Implications}
To validate finding robustness, all 520 final designs were independently re-evaluated by two additional evaluator models: \texttt{claude-opus-4-6} and \texttt{claude-sonnet-4-6}, using the same 12-dimension rubric as the primary evaluator (\texttt{openai/gpt-oss-120b-maas}). This yields 1,560 independent evaluations across three model families. The weighted ensemble ($2\times$Opus + $2\times$Sonnet + $1\times$GPT-OSS) is reported in Table~\ref{tab:rankings} above, and confirms all four core findings.

\subsection{Evaluator Agreement: What We Know for Certain}
All three evaluators agree on v4b structural adversarial supremacy. v4b ranks \#1 (GPT-OSS: 4.415, Opus: 4.553, Sonnet: 4.834). The structural adversarial advantage through rewrite mandates is the most robust finding in the entire experiment---all three evaluators place v4b at or very near the top, confirming that this topology achieves consistently high quality.

All three evaluators agree on merge failure. The v3 merge variants are bottom-tier for all evaluators (v3c: 3.734, 3.595, 3.675). Opus is harshest on naive merge (v3a: 3.443 vs. GPT-OSS 3.944), confirming the Frankenstein effect is universally recognized as a problem.

\subsection{Evaluator Disagreement: What Reveals Model Differences}
\begin{figure}
\centering
\includegraphics[width=\columnwidth]{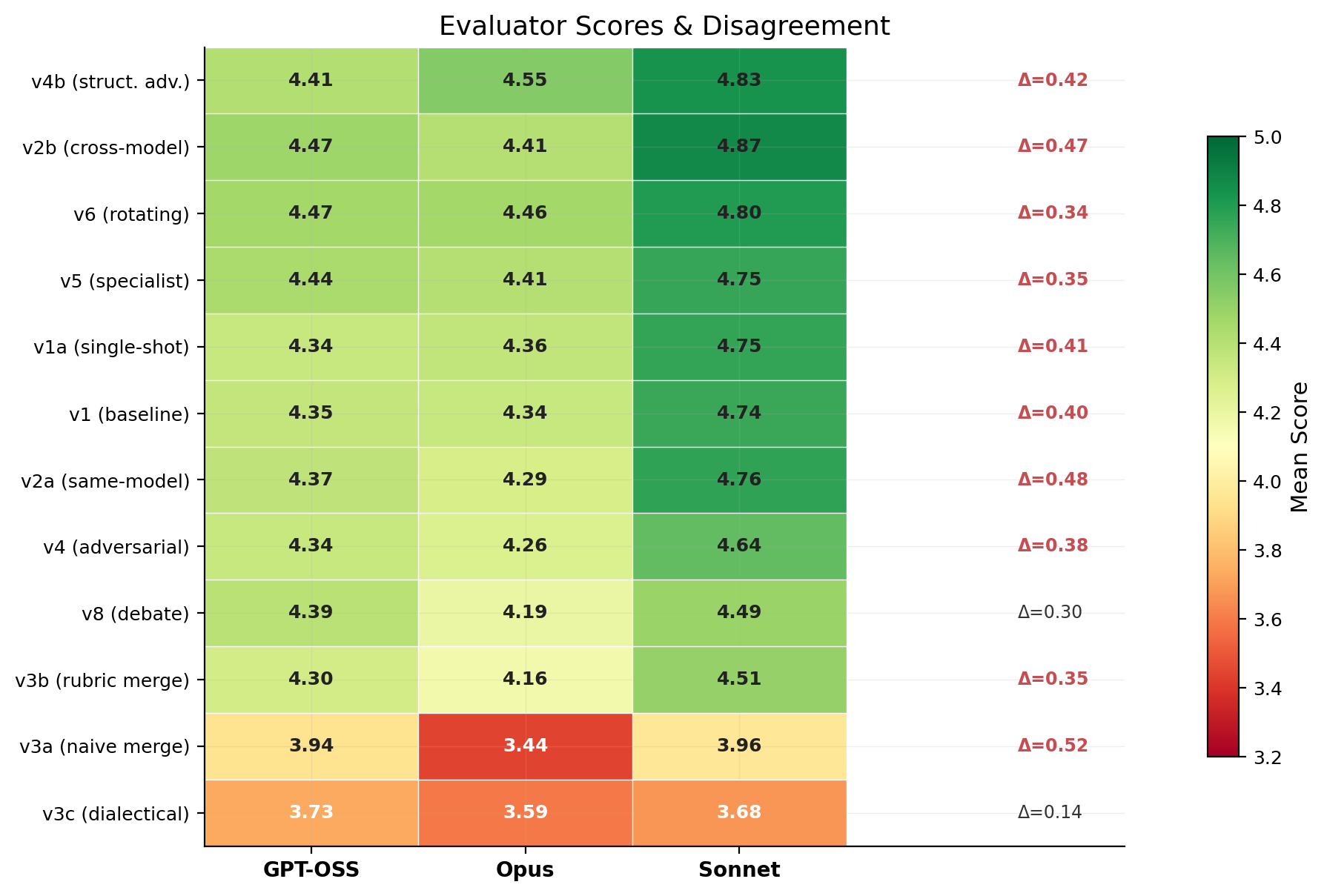}
\caption{Three-evaluator cross-validation: agreement (all rank v4b \#1, v3 bottom) and disagreement (v2b magnitude d=0.65 to d=1.44 reveals evaluator differences).}
\label{fig:crossval}
\end{figure}

The three evaluators diverge most sharply on v4b, the structural adversarial variant:
\begin{itemize}
\item \textbf{GPT-OSS 120B:} sees v4b improvement as small-medium ($\Delta$ = +0.070, d = 0.45 vs. v4)
\item \textbf{Claude Opus 4.6:} sees very large improvement ($\Delta$ = +0.289, d = 1.44)
\item \textbf{Claude Sonnet 4.6:} sees large improvement ($\Delta$ = +0.194, d = 1.11)
\end{itemize}
Both Claude models recognize more value in v4b's structural qualities---deeper architectural justification, mandatory rewrite sections, explicit trade-off analysis---than GPT-OSS does. This reflects systematic differences in how model families weight design rigor. The evaluator disagreement is itself a finding: it reveals that design quality is genuinely multifaceted and that different perspectives (different evaluator model families) highlight different dimensions of goodness. A single evaluator (whether GPT-OSS, Opus, or Sonnet) would produce incomplete or biased rankings; an ensemble is more robust. Fig.~\ref{fig:crossval} summarizes this three-evaluator cross-validation, highlighting both the points of agreement and the magnitude of disagreement.

\subsection{Implications for Methodology}
The three-evaluator validation has four key implications. First, the central finding---v4b structural adversarial is unanimously strong---is confirmed and robust. Second, core failures (v3 merge underperformance) are universally recognized. Third, evaluator disagreement is informative rather than problematic: where model families diverge on mechanisms like v2b cross-model review, it highlights genuine multifaceted quality and systematic differences in how evaluators weight architectural approaches. Fourth, a multi-evaluator ensemble produces more robust rankings than any single evaluator, just as multi-agent generation produces better designs than single-agent generation. The weighted ensemble methodology (prioritizing Claude's newer architectures while maintaining GPT-OSS signal diversity) is justified by this cross-validation evidence, especially in capturing both mechanisms (structural adversarial and cross-model review) that achieve excellence through different paths.

\end{document}